# Convective instability and boundary driven oscillations in a reaction-diffusion-advection model

Estefania Vidal-Henriquez, Vladimir Zykov, Eberhard Bodenschatz, and Azam Gholami

# Convective instability and boundary driven oscillations in a reaction-diffusion-advection model

Estefania Vidal-Henriquez,[1,a)] Vladimir Zykov,[1] Eberhard Bodenschatz,[1,2,3] and Azam Gholami[1,b)]
[1]*Max Planck Institute for Dynamics and Self-Organization, Am Fassberg 17, D-37077 Göttingen, Germany*
[2]*Institute for Nonlinear Dynamics, University of Göttingen, D-37073 Göttingen, Germany*
[3]*Laboratory of Atomic and Solid-State Physics and Sibley School of Mechanical and Aerospace Engineering, Cornell University, Ithaca, New York 14853, USA*

In a reaction-diffusion-advection system, with a convectively unstable regime, a perturbation creates a wave train that is advected downstream and eventually leaves the system. We show that the convective instability coexists with a local absolute instability when a fixed boundary condition upstream is imposed. This boundary induced instability acts as a continuous wave source, creating a local periodic excitation near the boundary, which initiates waves travelling both up and downstream. To confirm this, we performed analytical analysis and numerical simulations of a modified Martiel-Goldbeter reaction-diffusion model with the addition of an advection term. We provide a quantitative description of the wave packet appearing in the convectively unstable regime, which we found to be in excellent agreement with the numerical simulations. We characterize this new instability and show that in the limit of high advection speed, it is suppressed. This type of instability can be expected for reaction-diffusion systems that present both a convective instability and an excitable regime. In particular, it can be relevant to understand the signaling mechanism of the social amoeba *Dictyostelium discoideum* that may experience fluid flows in its natural habitat. © *2017 Author(s). All article content, except where otherwise noted, is licensed under a Creative Commons Attribution (CC BY) license ([http://creativecommons.org/licenses/by/4.0/](http://creativecommons.org/licenses/by/4.0/)).* https://doi.org/10.1063/1.4986153

In a reaction-diffusion-advection system, one or more species are carried away by a flowing medium with an externally imposed velocity. Therefore, the conditions of the system upstream become important to the phenomena observed downstream. In this work, we present the effects of adding an absorbing fixed boundary condition at the upstream end of the system. We focus on the convectively unstable regime, where a perturbation applied to the system dies out in the laboratory reference frame, while it grows in a moving one. By fixing the upstream boundary condition, the system becomes unstable, producing a trigger wave that travels upstream, and a wave train propagating downstream. The trigger wave is absorbed when it reaches the upstream boundary, then the system destabilizes again, and the phenomenon repeats. In 2-D simulations, the trigger wave propagating against the flow has a triangular shape, similar to the concentration profiles exhibiting a cusp in auto-catalytic advection reactions.[1,2] The here reported mechanism can be expected to be applicable to other reaction-diffusion-advection systems in order to produce a continuous, periodic influx of wave trains.

## I. INTRODUCTION

Many out of equilibrium phenomena in nature can be described by reaction-diffusion systems. This includes the Belousov-Zhabotinsky reaction,[3,4] electrical impulse dynamics in the heart,[5] skin patterns in fish,[6] calcium dynamics in oocytes,[7] and slime mold aggregation,[8] among others. In many cases, the active components of such reactions might be subjected to advective flows, which cause new kinds of instabilities.[9] The most commonly studied types of these instabilities are of convective or absolute nature.[10,11] Both types of instabilities have been observed in simulations,[9,12] as well as in experiments such as the Belousov-Zhabotinsky reaction.[13,14]

Due to the advective nature of the flow, the upstream boundary conditions have important consequences for the spatio-temporal dynamics downstream. Most studies have been performed with no-flux boundary conditions or periodic boundaries, which simplifies the analysis by going into a comoving reference frame. Under these boundaries, an initial perturbation creates a growing wave train[15,16] whose wavelengths and velocities depend on the particular characteristics of the system. However, the comoving frame analysis is impossible with a Dirichlet (fixed) boundary condition. In particular, an absorbing (zero amplitude) boundary condition corresponds to a one dimensional defect and is the one dimensional equivalent of a spiral center in excitable systems.[17–19] Up to now, the effects of this type of upstream

a)Electronic mail: estefania.vidal@ds.mpg.de
b)Electronic mail: azam.gholami@ds.mpg.de

condition on an advection-diffusion system have received little attention. Preliminary results on such a system were presented by Gholami *et al.*[20,21] where a continuous influx of wave trains was observed.

Here, we show that in the reaction-diffusion-advection system under study (see below), a particular kind of boundary induced instability occurs when the advection velocities are below a threshold. This boundary condition creates waves periodically with a period dependent on the imposed flow velocity. Unlike the commonly emitted waves by a boundary, these waves do not travel in just one direction (either towards or away from the boundary as is usual in these systems[22]), but instead two waves appear, one that travels towards and one that travels away from the boundary. In order for this to be possible, these waves do not grow directly at the boundary, but at a finite distance from it. The reaction of the system to this boundary driven instability is also different from the way it reacts to an external perturbation. In this system, a perturbation creates a growing wave train that is advected downstream, while in the absorbing boundary case, the growing instability produces not only a wave train downstream but also a wave travelling upstream.

The downstream wave train is equivalent to the one observed with the no-flux boundary condition. We fully characterized this wave train using linear stability analysis in a moving reference frame and calculated the periodic travelling wave solutions. The upstream travelling wave is the novel feature of this process. This wave travels upstream until it reaches the fixed boundary where it is absorbed, and the process starts again. This process creates wave trains with a period dependent on the imposed flow velocity and thus provides a mechanism to continuously generate wave trains in the fixed reference frame.

To investigate this effect, we performed numerical simulations in one dimension of a model proposed by Martiel and Goldbeter[23] which are reaction-diffusion equations, with the addition of an advection term due to an imposed external flow. To ensure accuracy in the simulations, we implemented a Runge-Kutta scheme with an adaptable time step based on the Merson error estimation.[24] To complete the study of the convectively unstable regime, we also performed linear stability analysis of the system in a moving reference frame and periodic travelling wave calculations which we compared with the full nonlinear system solutions. Finally, we performed numerical simulations in 2-Dimensions to study the effect of the flow profile on the boundary induced oscillations. Similar to fronts in advected auto-catalytic reactions,[1,2] we observed a strong triangular deformation of the trigger wave travelling upstream.

## II. THE REACTION-DIFFUSION-ADVECTION MODEL

Cellular slime moulds are unique organisms positioned between uni- and multi-cellular life in the evolutionary tree. The amoebae of the cellular slime mould *Dictyostelium discoideum* normally live as single cells in forest soil and feed on bacteria. They multiply by binary fission. Starvation induces a developmental program in which up to $10^5$ amoebae aggregate chemotactically to form a multicellular mass,[25] the so-called slug, that behaves as a single organism and migrates to search for food and better environmental conditions. On failing to find nutrients, the slug culminates into a fruiting body consisting of a stalk and a mass of spores.[26] Spores are dispersed by rain and small animals and under suitable conditions germinate to release amoebae and the whole cycle starts over again.

Cyclic adenosine monophosphate (cAMP) is the primary chemoattractant for the *D. discoideum* cells during early aggregation. cAMP is emitted from the aggregation centers in a pulsatile manner and surrounding cells detect it by highly specific cAMP receptors.[27] When cAMP binds to the receptors, it triggers a series of intracellular reactions that activate an enzyme called Adenylate cyclase (ACA), which in turns consumes Adenosine triphosphate (ATP) to produce intracellular cAMP. The cAMP produced inside the cell is partially degraded by intracellular phosphodiesterase and partially transported to the extracellular medium. Phosphodiesterase secreted by the cells degrades extracellular cAMP and suppresses the accumulation of excessive cAMP in the aggregation field (Fig. 1). Since each cell responds to cAMP by moving towards the source of cAMP, by emitting a pulse of cAMP itself, and by a refractory period, the cAMP signal is relayed outward from the aggregation center as a wave.[28,29] During the refractory period, the amoebas that have detected and produced cAMP do not react to it for a few minutes, and therefore, a new cAMP wave cannot pass during this period. This refractory phase is included in the model in terms of the membrane receptors. These receptors are present in two states, active and inactive. The first one has a higher probability to bind with cAMP than the second one. Once the receptors bind with cAMP, they change their state to inactive and then slowly change back to their active state. This combination of relay and refractory phase is characteristic of excitable systems and produces target patterns or spirals in two dimensional systems. The geometry of propagating waves is analogous to the spatio-temporal pattern of chemical waves in the Belousov-Zhabotinski reaction.[3,4]

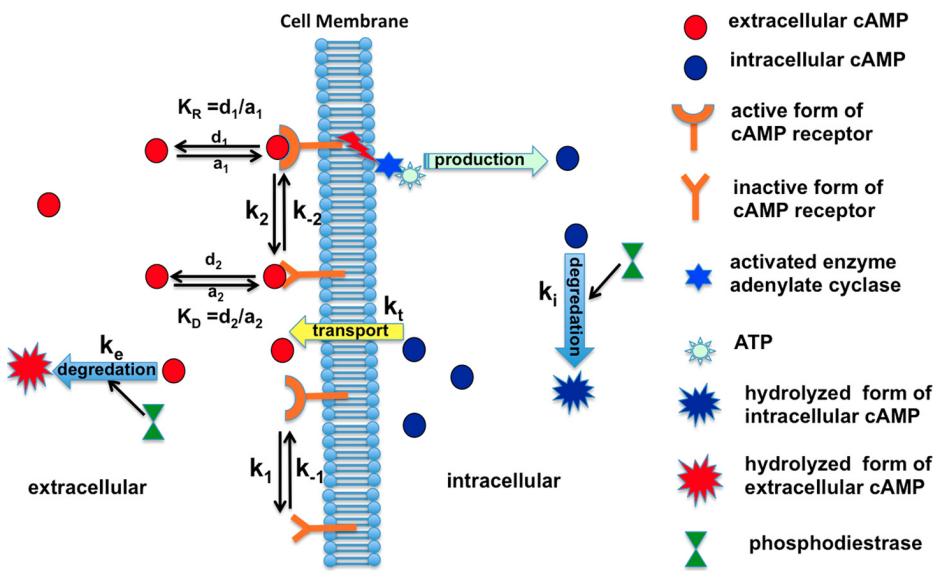

FIG. 1. Schematic representation of the model proposed by Martiel and Goldbeter for production and relay of cAMP in *D. discoideum*. The extracellular cAMP binds with the membrane receptors at a rate $K_R$ for those in the active state, activating ACA which in turn produces intracellular cAMP that is transported to the extracellular media at a rate $k_t$. After binding, the receptors change to their inactive form at a rate $k_1$ which has a lower probability of binding with cAMP ($K_D$). The extracellular cAMP is then degraded via phosphodiesterase at a rate $k_e$. Reproduced from Ref. 20.

The model we used for our study was initially proposed by Martiel and Goldbeter[23] and extended by Tyson *et al.*[30] In this reaction-diffusion system, the concentration of the signalling chemical cAMP is the activator, while the cAMP receptors on the cell membrane act as inhibitors. Since the inhibitor is cell bounded and we assume that the imposed flow is not strong enough to detach the cells from the substrate,[31] we add the advection term only to the activator dynamics. A detailed model derivation and the biological correspondence of the model parameters can be found in Ref. 23.

The main equations of the model in its three component version are as follows, where $\rho$ stands for the percentage of active receptors on the cell membrane, $\gamma$, the extracellular concentration of cAMP, and $\beta$, the intracellular amount of cAMP. The receptor dynamics are given by

$$\partial_t \rho = -k_1 f_1(\gamma)\rho + k_1 f_2(\gamma)(1-\rho)$$

with

$$f_1(\gamma) = \frac{1+\kappa\gamma}{1+\gamma}, \quad f_2(\gamma) = \frac{\mathcal{L}_1 + \kappa\mathcal{L}_2 c\gamma}{1+c\gamma},$$

where $f_1$ controls the receptor desensitization (change from active to inactive state) and $f_2$, the resensitization. The intracellular cAMP is increased by the cAMP production, which in turn depends on the extracellular cAMP and the active receptors. This production is tuned through the rate $\sigma$ at which the activated ACA produces cAMP. The intracellular cAMP is diminished through degradation by intracellular phosphodiesterase at a rate $k_i$ and passive transport outside of the cell at a rate $k_t$

$$\partial_t \beta = q\sigma\alpha\Phi(\rho,\gamma)/(1+\alpha) - (k_i + k_t)\beta$$

with

$$\Phi(\rho,\gamma) = \frac{\lambda_1 + Y^2}{\lambda_2 + Y^2}, \quad Y = \frac{\rho\gamma}{1+\gamma},$$

$\lambda_2 = (1+\alpha\theta)/(\epsilon(1+\alpha))$, and $\lambda_1 = \lambda\theta/\epsilon$. The extracellular concentration of cAMP $\gamma$ is degraded at a rate $k_e$ by the extracellular phosphodiesterase and is increased by the transport of cAMP from the intracellular medium

$$\partial_t \gamma = D\nabla^2\gamma - v\cdot\nabla\gamma + k_t\beta/h - k_e\gamma.$$

We nondimensionalize the system by introducing dimensionless time and space as $t' = t\cdot k_1$ and $x' = x\cdot k_1/\sqrt{k_e D}$. Dropping primes and setting $\epsilon_1 = k_1/k_e$, $\epsilon' = k_1/(k_i + k_t)$, we arrive at

$$\partial_t \rho = -f_1(\gamma)\rho + f_2(\gamma)(1-\rho), \quad (1a)$$

$$\epsilon'\partial_t \beta = q\sigma\alpha\Phi(\rho,\gamma)/((1+\alpha)(k_i+k_t)) - \beta, \quad (1b)$$

$$\partial_t \gamma = \epsilon_1\nabla^2\gamma - v\cdot\nabla\gamma + (k_t\beta/(hk_e) - \gamma)/\epsilon_1. \quad (1c)$$

Finally, we reduce this system to a two component model which simplifies its theoretical treatment. For this, we assume $\epsilon'$ small, which means that the intracellular cAMP is instantaneously transmitted to the outside media (for a discussion on the validity of this approximation, refer to Refs. 23 and 30). We then arrive at the two component Martiel-Goldbeter, which we will use during the rest of this paper

$$\partial_t \gamma = \epsilon_1 \nabla^2\gamma - v\cdot\nabla\gamma + (s\Phi(\rho,\gamma) - \gamma)/\epsilon_1, \quad (2a)$$

$$\partial_t \rho = -f_1(\gamma)\rho + f_2(\gamma)(1-\rho), \quad (2b)$$

where $s = qk_t\alpha\sigma/(k_e(k_t+k_i)h(1+\alpha))$. All used parameters are listed in Table I and were selected as suggested by Lauzeral *et al.*[32] because of their good agreement with experimental measurements. We selected $\sigma$ and $k_e$ as control parameters since they account for the production and degradation of extracellular cAMP, respectively. Depending on these two parameters, this system can have one, two, or three steady state solutions, as is shown in the phase diagram in Fig. 2. We focused on the range where only one steady state exists (green, yellow, and blue in Fig. 2). We performed linear stability analysis around this steady state solution $(\gamma_0, \rho_0)$ by setting $\gamma = \gamma_0 + \gamma'$, $\rho = \rho_0 + \rho'$, linearizing, dropping primes, and performing Fourier transform

$$(\gamma, \rho) = \int_{-\infty}^{\infty}(\gamma_k, \rho_k)e^{\omega(k)t + ikx}dk,$$

we arrive at the dispersion relation

$$0 = \omega^2 + \omega(-T + \epsilon_1 k^2 + ivk) + \Delta - a_{22}(\epsilon_1 k^2 + ivk), \quad (3)$$

where $\Delta = a_{11}a_{22} - a_{12}a_{21}$, $T = a_{11} + a_{22}$,

TABLE I. Parameters used for simulations of Eq. (2).

| | | |
|---|---|---|
| $c = 10$ | $h = 5$ | $k_1 = 0.09\,\text{min}^{-1}$ |
| $k_2 = 1.665\,\text{min}^{-1}$ | $K_R = 10^{-7}\,\text{M}$ | $k_i = 1.7\,\text{min}^{-1}$ |
| $k_t = 0.9\,\text{min}^{-1}$ | $\mathcal{L}_1 = 10$ | $\mathcal{L}_2 = 0.005$ |
| $q = 4000$ | $\epsilon = 1$ | $\lambda = 0.01$ |
| $\theta = 0.01$ | $\alpha = 3$ | $D = 0.024\,\text{mm}^2\cdot\text{min}^{-1}$ |

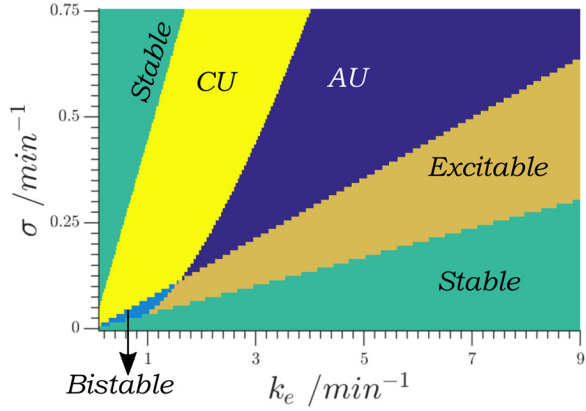

FIG. 2. Phase diagram of the system described by Eq. (2). Stable regime in green, in this regime only one solution exists and is stable. In the yellow area labeled CU exists one steady state that is convectively unstable. In the AU labeled blue area exists one unstable steady state surrounded by a limit cycle. The orange regime marked as excitable presents three steady states, one of which is excitable, while the light blue bistable regime has three steady states, two of which are stable.

$$a_{11} = \frac{s}{\epsilon_1} \frac{2\rho_0^2 \gamma_0 (\lambda_2 - \lambda_1)}{(1+\gamma_0)^3 (\lambda_2 + Y_0^2)^2} - \frac{1}{\epsilon_1},$$

$$a_{12} = \frac{s}{\epsilon_1} \frac{2\gamma_0^2 \rho_0 (\lambda_2 - \lambda_1)}{(1+\gamma_0)^2 (\lambda_2 + Y_0^2)^2},$$

$$a_{21} = (1-\rho_0) \frac{\kappa \mathcal{L}_2 c - c \mathcal{L}_1}{(1+c\gamma_0)^2} - \frac{\rho_0(\kappa - 1)}{(1+\gamma_0)^2}, \quad \text{and}$$

$$a_{22} = -f_1(\gamma_0) - f_2(\gamma_0).$$

From here, the different regimes can be distinguished. Starting with the left green area of the phase diagram of Fig. 2 and with no imposed flow velocity $v=0$, the system has $T<0$ and $\Delta >0$; therefore, $Re(\omega) < 0$ for every $k$ and the system is stable. Increasing $k_e$, the system has a Hopf bifurcation (at the boundary between yellow and blue area in Fig. 2) and a limit cycle appears (Oscillatory regime). When $v \neq 0$, part of the stable regime becomes convectively unstable (yellow in Fig. 2). In this area, $a_{11}$ is positive and we can calculate the minimum imposed velocity at which the system becomes unstable, by calculating when the real part of $\omega$ becomes positive. This gives the following relation:

$$v^2(k) = \frac{(-\Delta/a_{22} + \epsilon_1 k^2)(\epsilon_1 k^2 - T)^2}{k^2(a_{11} - \epsilon_1 k^2)}. \quad (4)$$

This is a convex curve dependent on $k$ with asymptotes at $k=0$ and $k=\sqrt{a_{11}/\epsilon_1}$. Its global minimum corresponds to the critical velocity $v_c$ at which the system destabilizes. This type of instability is of the convective type, which means that although a perturbation applied to the system will die out in the laboratory reference frame, it will grow in a reference frame moving with a speed $v'$, when the system is advected with a flow higher than $v_c$. All our simulations were performed in this regime.

Before proceeding to the characterization of the boundary driven instability, we perform a general description of the wave trains present in this system.

## III. NO-FLUX BOUNDARY CONDITION

In the convectively unstable regime, when the advection velocity $v$ is above the critical value $v_c$ [calculated as the minimum of Eq. (4)], a perturbation creates a peak that is advected downstream. This peak creates further peaks behind it, producing a wave train, as can be observed in Fig. 3. The front of this wave train travels with a speed $v_f$ higher than the imposed flow $v$, while the rear of the wave train travels with a velocity $v_b < v$. This difference between $v_b$ and $v_f$ translates into the wave train growing in size and having more peaks as time passes. These velocities are indicated by colored lines in Fig. 3. The characteristics of these wave trains can be estimated by taking the Fourier transform in a moving reference frame $y = x - v't$, where $v'$ is a free parameter

$$(\gamma, \rho) = \int_{-\infty}^{\infty} (\gamma_k, \rho_k) e^{t(\omega + ikv') + iky} dk,$$

with $k \in \mathbb{C}$ and $\omega(k)$ given by the dispersion relation, Eq. (3). According to the method of steepest descents,[10] the long term behavior of this integral is given by the saddle point of the term accompanying $t$, i.e.,

$$\frac{d}{dk}(\omega(k) + ikv') = 0.$$

Since $\omega$ is also complex, we can use the Cauchy-Riemann Equations

$$\frac{\partial \omega_r}{\partial k_r} = \frac{\partial \omega_i}{\partial k_i} = 0 \quad \text{and} \quad \frac{\partial \omega_r}{\partial k_i} - v' = \frac{\partial \omega_i}{\partial k_r} + v' = 0, \quad (5)$$

where $k = k_r + ik_i$ and $\omega = \omega_r + i\omega_i$. This gives pairs of solutions $(k, v')$, each with its growing rate $\lambda_r = \omega_r - k_i v'$. A typical curve $\lambda_r$ vs $v'$ is shown in Fig. 4. The maximum of this curve corresponds to the group velocity of the wave train, it is the fastest growing mode and has $k_i = 0$, $k_r \neq 0$. To calculate the edges of the wave train, the relevant values are the pairs with zero growing rate, because these will correspond to the first and last points at which the system destabilizes and therefore mark the boundaries of the velocity range at which the wave train can be observed. There are two velocities $v'$ with zero growing rate, the lower corresponds to $v_b$ and the higher to $v_f$. This linear calculation has a very good agreement with the velocities calculated from the numerical simulations of the full nonlinear system, Eq. (2). This is shown in Fig. 5 where these two data sets are compared.

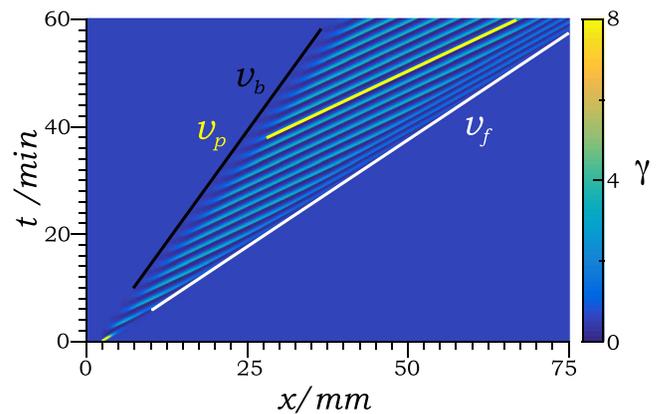

FIG. 3. Space-time plot of a simulation performed in the convectively unstable regime using no-flux (Neumann) boundary condition. The wave train is generated by an initial perturbation and is advected downstream (to the right) by the imposed flow. The relevant velocities present in the wave train are highlighted, these are the velocity of wave train rear $v_b$ in black, front velocity $v_f$ in white, and individual peak velocity $v_p$ in yellow. All numerical simulations were performed using $k_e = 3.0 \text{ min}^{-1}$ and $\sigma = 0.45 \text{ min}^{-1}$.

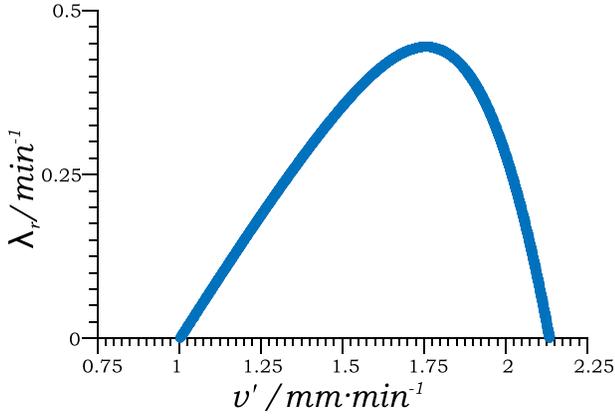

FIG. 4. Growth rate in the different reference systems $v'$ for $v = 2$ mm/min. The intersections with the x-axis mark the back and front velocities of the wave train. For these parameters $v_b = 1.00$ mm/min and $v_f = 2.13$ mm/min.

It has been shown that for some systems, the previously used method may not catch the fastest growing mode in a moving reference frame.[33,34] For this, the more reliable Briggs collision criterion[35] is recommended. However, in our system the function $\omega(k)$ has only two local maxima and one unstable branch for real $k$, and under these conditions, the saddle point approach is enough to find all the unstable points.[36]

To connect to other results in literature, it is worth mentioning that in our calculation, $v_f$ is equivalent to the spreading speed to the right of the system,[37,38] which means that it is the supreme of the velocities $v'$, such that the system is unstable in the comoving frame moving at $v'$.

To characterize each individual peak velocity $v_p$, we studied the periodic travelling wave solutions of this system. These waves are characteristic of oscillatory systems[39] and have the property $\gamma(z + T) = \gamma(z)$ with $z = x - ct$ for a certain combination of propagation velocity $c$ and period $T$. The wave calculation and stability analysis were performed using the software Wavetrain.[40–42]

We found a range of velocities $c$ at which the periodic travelling wave solutions exist. Inside this range, there is a band of velocities $c$ where they are stable. The velocities of each individual peak fall into this band as shown in Fig. 6. The selection of a particular wave solution depends on the initial conditions.

The velocity of each particular peak $v_p$ is higher than the front velocity; therefore, each peak moves forward in the train until it approaches the front, where it has to slow down until it matches $v_f$, the velocity of the front of the wave train. Since wavelength and velocity are uniquely linked, the peaks closer to the front of the wave train have a smaller wavelength than the rest of the train. This creates a traffic jam where more peaks start to accumulate in this shorter wavelength area at the front of the train. A similar process has been observed in other reaction-diffusion systems.[43,44]

## IV. FIXED UPSTREAM BOUNDARY CONDITION

We performed numerical simulations with a Dirichlet (fixed) boundary condition upstream $\gamma(x = 0) = \rho(x = 0) = 0$ in the convectively unstable regime. We found that for very high flow speeds, the advection dominates over the diffusion and the system reaches a stable extended steady state. This state can be approximated in powers of $\delta = \epsilon_1/v^2$ with the time independent version of Eq. (2) as

$$\delta \partial_{x'x'}\gamma = \partial_{x'}\gamma - (s\Phi(\rho,\gamma) - \gamma)/\epsilon_1,$$
$$\rho = f_2(\gamma)/(f_1(\gamma) + f_2(\gamma)),$$

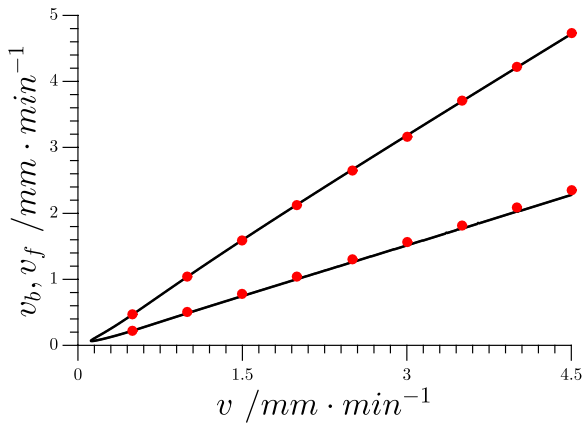

FIG. 5. Dependence of the wave train velocities on the imposed advection flow. The lower value corresponds to the back of the train, that is, the first point that destabilises, while the higher value corresponds to the front of the wave train, that is, the last point that destabilises. The continuous line corresponds to the prediction obtained by the linear analysis, and the dots are the values obtained from the simulations of the full nonlinear equations.

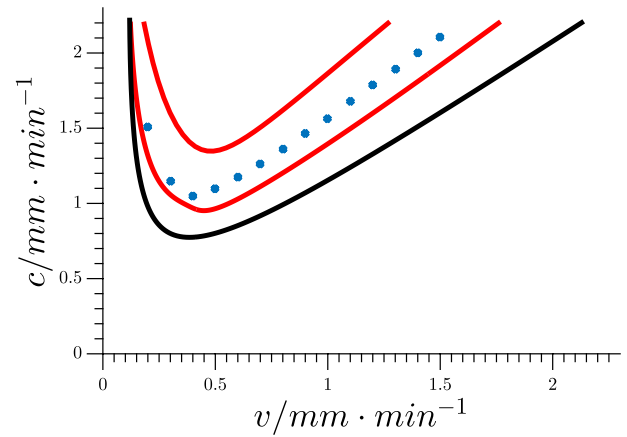

FIG. 6. Region of existence of periodic travelling wave solutions. $v$ stands for the imposed advection velocity, and $c$, the velocity of the periodic travelling wave. The solutions exist above the black line and are stable in the band between the red lines. The dots correspond to the solution selected by the system in the middle of the wave train in our numerical simulations.

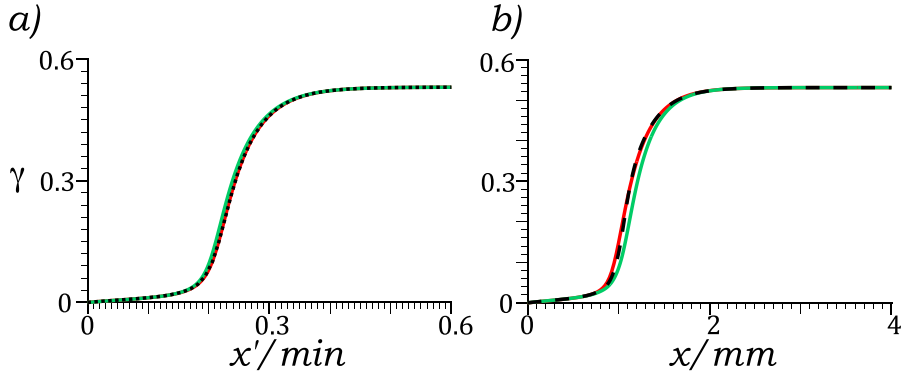

FIG. 7. (a) High speed solution with Dirichlet boundary condition. Advection velocity $v = 2$ mm/min in green and $v = 5$ mm/min in red. Zero order approximation $\varphi_0$ solution of Eq. (6a) in black dotted lined. Scaled space $x' = x/v$. (b) Comparison for approximation at smaller speed. Full solution with Dirichlet boundary condition and $v = 1.33$ mm/min in red. Zero order approximation $\varphi_0$ in green and first order approximation $\gamma = \varphi_0 + \delta\varphi_1$ in dashed black line.

where $x' = x/v$ and $\gamma(x' = 0) = 0$. The first two terms of the expansion were calculated taking $\gamma = \varphi_0 + \delta\varphi_1$

$$0 = \partial_{x'}\varphi_0 - (s\Phi(\varphi_0) - \varphi_0)/\epsilon_1, \quad (6a)$$

$$\partial_{x'x'}\varphi_0 = \partial_{x'}\varphi_1 - \varphi_1\left(s\frac{d\Phi}{d\gamma}\bigg|_{\gamma=\varphi_0} - 1\right)/\epsilon_1. \quad (6b)$$

This solution connects smoothly the zero boundary condition with the steady state of the system. This approximation matches quite well with the full solution as it is shown in Fig. 7.

We performed numerical linear analysis of this monotone solution and found that it becomes unstable at smaller velocities (when $\delta$ gets larger). The eigenvalues cross the real axis with non-zero imaginary part when the imposed flow velocity $v$ is lowered below a threshold. This bifurcation is shown in Fig. 8. The fastest growing eigenvector has the shape of a peak centered close to the fixed border, the distance between the peak and the border increasing with increasing imposed flow velocity.

To study this instability, we performed numerical simulations with Dirichlet boundary condition upstream and small imposed flow velocities. We observed that the system initially reaches a state similar to the one showed in Fig. 7, that is, a smooth connection between the boundary and the steady state. However, this solution becomes unstable producing a peak which, as it grows, divides into two peaks. One of the peaks travels downstream and produces a wave train as was previously described in Sec. III. The second peak travels upstream until it reaches the boundary. Once the upstream travelling peak has been absorbed by the boundary, the system goes back to the smooth solution, which then again becomes unstable and repeats the cycle. This whole process generates periodically wave trains propagating downstream, as shown in Fig. 9.

The period of these perturbations is hard to measure downstream due to the wave train that it generates, whose period is given by the periodic travelling wave solution. To solve this, we measured the period of the initial destabilization peak at its point of creation, as shown in white in Fig. 9. This nucleation location moves farther away from the boundary as the imposed flow velocity increases. This relation is shown in Fig. 10.

This period $T$ does not appear to have a relation to any of the periods in the train wave previously studied. This, combined with the difference in the back and front velocities $v_b$ and $v_f$, produces phase slips. The phase slips occur when the front of the newly generated wave train catches up with the back of the previous wave train, thus forming downstream

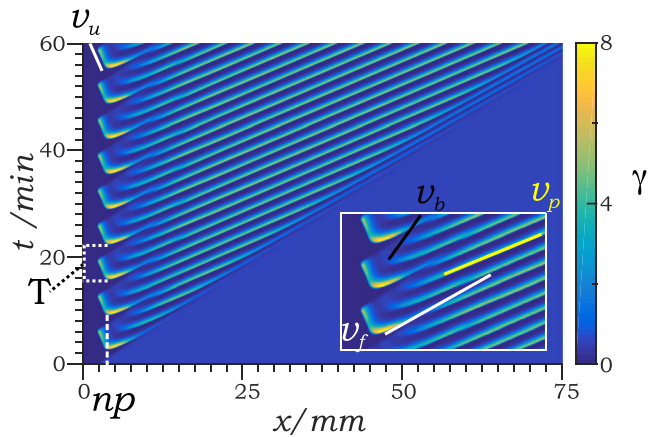

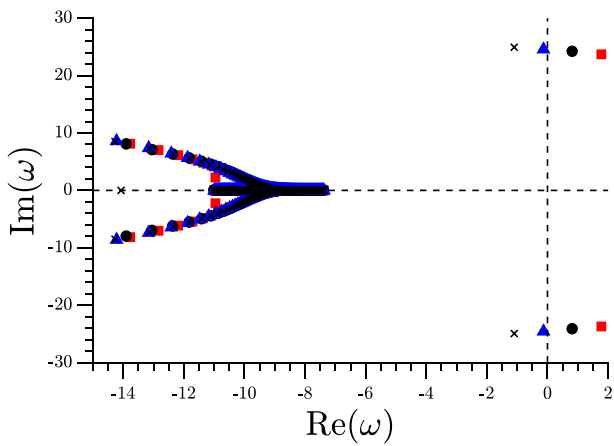

FIG. 8. Frequencies of the linear analysis of the monotone profile $\gamma = \varphi_0 + \delta\varphi_1$ showing the oscillatory bifurcation when the imposed flow $v$ is lowered. $v = 1.32$ mm/min in red squares, $v = 1.33$ mm/min in black circles, $v = 1.34$ mm/min in blue triangles, and $v = 1.35$ mm/min in black crosses.

FIG. 9. Space-time plot of simulation performed in the convectively unstable regime using fixed (Dirichlet) boundary condition. Wave trains generated by the instability described in Sec. IV and measured quantities highlighted. The nucleation point is $np$, where the destabilization occurs, $T$, the oscillations period, and $v_u$, the velocity of the upstream travelling peak. Inset with a zoom of the wave generation area with previously defined quantities of the wave train highlighted, $v_b$, velocity of the back of the wave train, $v_f$, velocity of the front of the wave train, and $v_p$, velocity of each individual peak. The imposed flow velocity is 1.2 mm/min.

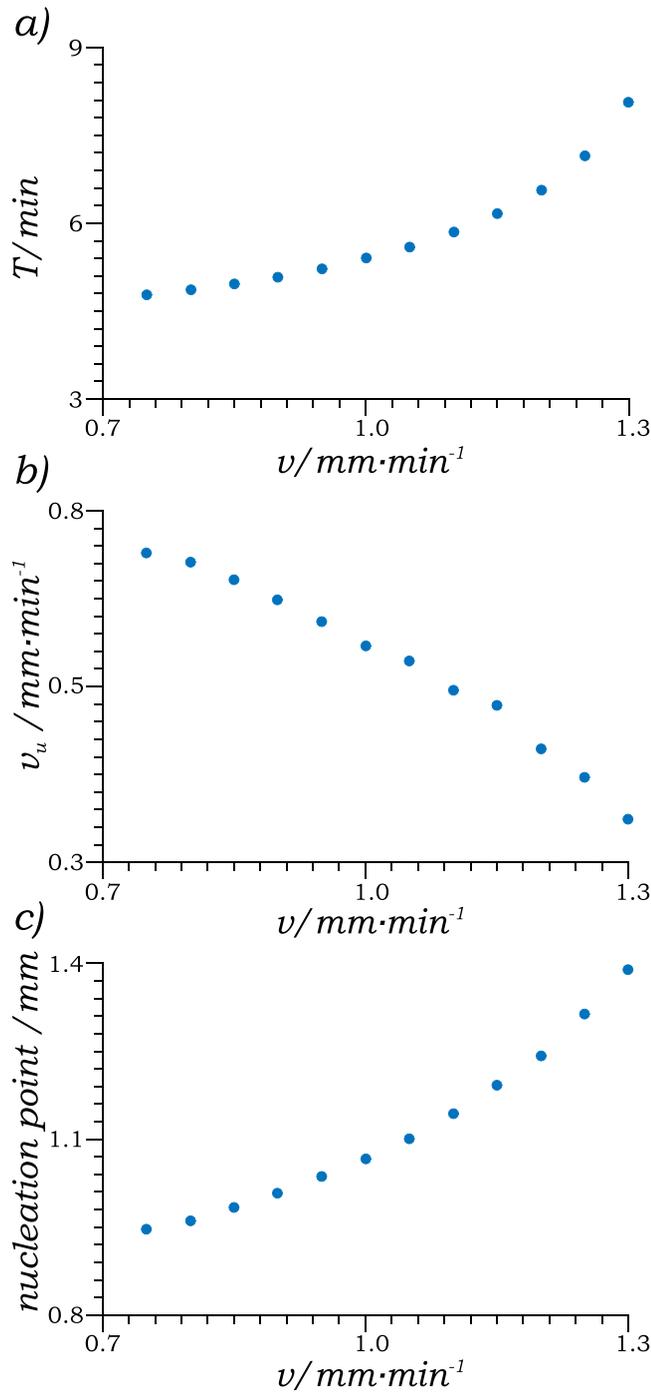

FIG. 10. Different properties of the boundary driven oscillations as function of the imposed flow velocity $v$. (a) Period of the oscillations, as measured at the point of nucleation of the instability. (b) Velocity of the upstream travelling peak. (c) Destabilization point position.

one larger wave train with phase slips. This process is highlighted on the inset of Fig. 9.

As expected, the velocity of the upstream travelling peak $v_u$ decreases with imposed flow velocity. Since the new wave does not appear until the previous one has travelled up to the boundary, the instability period is directly affected by the velocity of the upstream peak. Therefore, the period increases with increasing imposed flow velocities. All these dependencies are shown in Fig. 10.

The periodical travelling wave solutions selected by the boundary condition could not be measured for every velocity value, because of the interaction between the new wave train and the old one. This interaction produces numerous phase slips that change the wavelength along the wave train. For those values of $v$ where it was measured, the selected travelling wave falls into the stable range shown in Fig. 6.

We understand the upstream travelling peak as a trigger wave, analogous to the ones present in the excitable regime in this system. Trigger waves are non-linear excitation waves that propagate in excitable media when a perturbation above a threshold is applied. In these systems, small perturbations damp out but supra-threshold ones are amplified and excite the neighboring area allowing for wave propagation.[45] A trigger wave has a velocity which is nonlinearly selected by the system. Another important characteristic is that a new trigger wave cannot enter the system until some recovery time has elapsed. In the case of the upstream travelling wave, the system cannot sustain another upstream travelling peak until the old one has reached the boundary and the cAMP close to the boundary has been washed away. Schematically, the wave works as follows. The cells closer to the boundary have been exposed to very small amounts of cAMP because it is initially washed away due to the boundary. As a result, they have a very high percentage of active receptors on the cell membrane. Therefore, they quickly react to the small perturbation of cAMP produced by the growing peak, emitting cAMP themselves and producing a trigger wave. It has been shown that trigger waves can travel against imposed flows when the advection is not too strong, experimentally in the Belousov-Zhabotinsky reaction[46] and numerically in the excitable regime of the Martiel-Goldbeter model[47] and in the FitzHugh-Nagumo model.[48]

## V. 2-DIMENSIONAL RESULTS

To study the instability already investigated in one dimension, we performed numerical simulations in a 2-Dimensional system. The dimensions were chosen following the *D. discoideum* experiments of Gholami *et al.*[31] In this microfluidic setup, the amoebas were placed in a 30 mm × 2 mm × 100 $\mu$m channel, where a constant flow was applied along the longest axis. Because of the small height and velocities of this system, the flow can be assumed to be laminar and constant in the long channel axis ($x$-axis), thus making a Poiseuille flow. We solved the Navier-Stokes equation under these assumptions and used this flow as our imposed advection for the simulations. The resulting flow is parabolic in the short axis ($z$-axis). This is the direction over which we averaged to have a 2-Dimensional system. In the $xy$-plane, the flow is almost planar in the center with a sharp boundary layer of the order of 50 $\mu$m on the top ($y = 2$ mm) and on the bottom ($y = 0$ mm) boundaries, where the velocity quickly drops to zero. The detailed flow profile calculation is presented in the Appendix.

We performed numerical simulations with no-flux boundary conditions on the top and bottom boundaries and Dirichlet $[\rho, \gamma(x = 0) = 0]$ boundary condition upstream. The simulations confirmed our previous observations in one dimension: When a small advection flow is applied, an instability appears, which creates a wave train downstream and a travelling peak upstream. This process can be observed in Fig. 11, the destabilization peak begins to appear in Fig. 11(b), creating a train wave. The back travelling wave is already visible in Fig. 11(d) and more clear in Fig. 11(f).

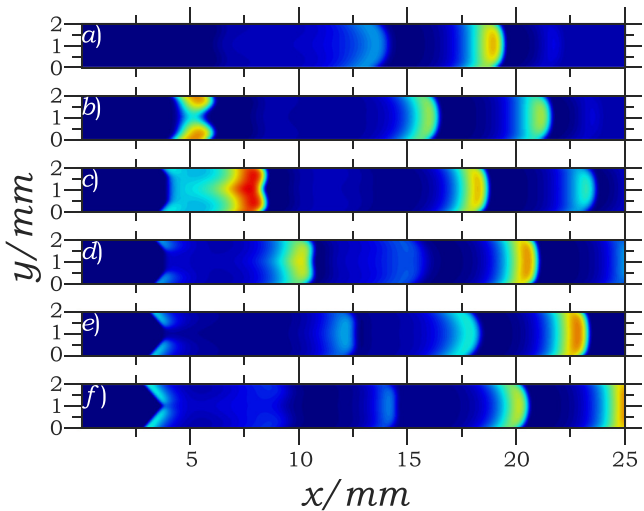

FIG. 11. Colormap of the concentration $\gamma$ every 0.9 min starting at $t = 10.5$ min at the top and increasing towards the bottom. Applied flow is $v = 1.75$ mm/min.

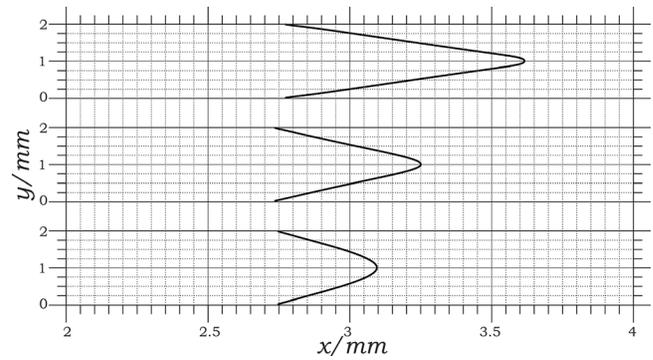

FIG. 13. Shape of the upstream travelling peak for different velocities taken as a contour at $\gamma = 1.0$. Top: $v = 1.75$ mm/min, middle: $v = 1.00$ mm/min, bottom: $v = 0.80$ mm/min. Different scales are used in the *x*- and *y*- axis for better visualisation.

Remarkable in comparison with the 1-Dimensional simulations are the range of existence of the instability and the shape of the upstream travelling peak. In the 2-Dimensional simulations, we observed that the system becomes stable at a higher speed $v = 1.75$ mm/min compared to the 1-Dimensional ones $v = 1.33$ mm/min, when measured at the center of the channel ($\sigma = 0.45$ and $k_e = 3.0$). We attribute this difference to the smaller advection speeds at the boundary layer which are enough to destabilize the whole system. This phenomenon was also observed in some preliminary simulations using a parabolic advection flow,[21] where the advection flow velocity is much smaller in a wider region, thus making the instability range of existence much larger.

Of particular interest is the shape that the upstream travelling peak acquires while it travels towards the boundary. Since this peak travels against the flow, its shape gets deformed due to the different speeds along the perpendicular axis. When the peak originally appears, it has a much flatter shape, similar to the imposed flow, as can be observed at the far right of Fig. 12. As the peak travels upstream (towards the left), it gets increasingly deformed until it acquires a triangular shape. Contours of the peak taken every 0.5 min are displayed in Fig. 12 showing this process.

The triangular deformation of a front due to an adverse flow was theoretically predicted by Edwards[1] and experimentally confirmed by Leconte *et al.*[2] for an auto-catalytic reaction. The main difference with our system is that in our reaction-diffusion-advection system, only the activator $\gamma$ is advected, while the inhibitor $\rho$ remains static. Like in those systems, the deformation of the wave is larger at larger imposed flows. This is shown in Fig. 13 for three different advection velocities. This wave deformation makes the characterization of the system difficult, because it produces different arrival times at the boundary. More work is needed in this direction to fully characterize this system in 2-D.

## VI. CONCLUSIONS

### A. No-flux boundary

We have analyzed and characterized the convectively unstable regime in the model proposed by Martiel and Goldbeter for cAMP production in *D. discoideum*. In this regime, an initial perturbation generates a wave train of growing size (i.e., it contains more peaks as time passes) that travels downstream. In particular, the speed of the peaks located near the wave front (back) is higher (lower) than the advection flow, thus causing the growing size. These two velocities were numerically characterized through linear stability analysis and have an excellent agreement with the velocities measured in the nonlinear simulations of the model. The growing mode on the center of the wave train corresponds to one of the periodic travelling wave solutions of the system and moves faster than the front of the train. Therefore, a peak will move towards the front of the train, where then it will decrease its speed to match the front velocity. As a result of this smaller speed, the wavelength near the front of the train is smaller than in the center of the wave train, thus producing a traffic jam. These wave trains are similar to the differential flow induced convective instability (DIFICI) waves which were first predicted in Ref. 9, experimentally observed in Ref. 13 and further investigated in Refs. 14–16.

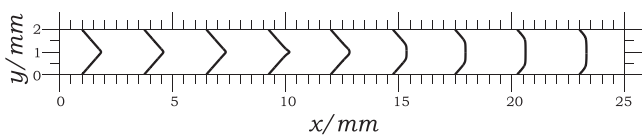

FIG. 12. Shape of the upstream travelling peak at different times taken as a contour at $\gamma = 0.4$ and $v = 1.75$ mm/min. Lines taken every 0.5 min, starting at $t = 12$ min on the right until $t = 16$ min on the left. Contours not in their original positions but spatially separated for better visualization.

## B. Fixed boundary

When slow advection speeds are applied along with a Dirichlet (absorbing) boundary condition, an instability appears that periodically produces wave trains. This instability initially generates a peak that divides into two, with one peak travelling upstream towards the boundary and the other one producing a wave train downstream. Once the peak travelling upstream has reached the absorbing boundary, the process starts again, thus acting as a continuous source of waves. The velocity of the wave travelling upstream is affected by the imposed flow velocity. As expected, it travels slower at higher advection, and since the instability does not appear until the peak reaches the boundary, this affects the period of the oscillation. The faster the imposed flow, the longer the period. The location of appearance of this instability also increases with the advected flow velocity. We emphasize that these upstream travelling waves are nonlinearly selected and depend solely on the system parameters.

This instability was also observed in 2-Dimensional simulations, where the upstream travelling peak acquires the triangular shape of fronts propagating against adverse flows.[1] This triangular shape increases its height with increasing advection flow. The instability persists up until higher velocities than in one dimension and similarly increases period with increased imposed flow.

The observed phenomena is different from other wave trains emitted by Dirichlet boundary conditions[22] in that the waves are not directly emitted or absorbed by the boundary, but instead appear as a pair of waves from a nucleation point which exists downstream from the boundary. From this pair of waves, one travels upstream and is absorbed by the boundary while the other creates a wave train downstream. We expect that a similar mechanism may exist in systems where the convective or absolute unstable regime exists close to an excitable regime, thus facilitating the creation of an upstream travelling peak. This mechanism can then be used to produce a constant wave influx.

## ACKNOWLEDGMENTS

The authors acknowledge A. Bae for fruitful discussions and unknown referees for insightful comments. E.V.H. thanks the Deutsche Akademische Austauschdienst (DAAD), Research Grants—Doctoral Programs in Germany. A.G. acknowledges MaxSynBio Consortium, which is jointly funded by the Federal Ministry of Education and Research of Germany and the Max Planck Society.

## APPENDIX: POSEUILLE FLOW CALCULATION

To estimate the flow profile inside the channel, we used the Navier-Stokes equations and assumed incompressible flow in a 3D rectangular geometry ($x \in [0, L], y \in [-c, c], z \in [-b, b]$), with zero velocity as the boundary condition along the two shortest directions, $\boldsymbol{u}(y = \pm c) = 0$ and $\boldsymbol{u}(z = \pm b) = 0$, thus obtaining

$$\rho \frac{D\boldsymbol{u}}{Dt} = \rho \boldsymbol{g} - \nabla p + \mu \nabla^2 \boldsymbol{u},$$

where bold text denotes vectors, $\mu$ is the system viscosity, $\rho$ its density, $p$ the pressure, and $\boldsymbol{u}$ the fluid velocity. We further simplified by assuming that the flow is constant over time, only exists in the $\hat{x}$-direction, and is constant over this direction, that is, $\boldsymbol{u} = u\hat{x}$ and $\partial_x u = \partial_t u = 0$; therefore, the previous equation reduces to

$$\mu \left( \frac{\partial^2 u}{\partial y^2} + \frac{\partial^2 u}{\partial z^2} \right) = \frac{\partial p}{\partial x} \equiv -G,$$

where $G$ is an externally applied pressure difference. This can be solved by setting the auxiliary function

$$\mathcal{F} = u - \frac{G(b^2 - z^2)}{2\mu},$$

which reduces the system to solve $\nabla^2 \mathcal{F} = 0$ with boundary conditions $\mathcal{F}(z = \pm b) = 0$ and $\mathcal{F}(y = \pm c) = -G(b^2 - z^2)/2\mu$. Using variable separation $\mathcal{F}(y, z) = F_y(y)F_z(z)$ and considering the symmetry of the system, we obtain $F_z = \cos(k_z z)$, $F_y = \cosh(k_y y)$ with $k_z = k_y$. The boundary condition $F_z(z = \pm b) = 0$ sets

$$k_z = k_y = \frac{m\pi}{2b}$$

with $m$ an odd integer. The other boundary condition is fulfilled by using Fourier series, finally obtaining

$$u = \frac{G(b^2 - z^2)}{2\mu} + \sum_{odd} A_n \cosh\left(\frac{n\pi}{2b}y\right)\cos\left(\frac{n\pi}{2b}z\right),$$

where

$$A_n = -\frac{16Gb^2}{\mu\pi^3 n^3} \frac{\sin\left(\frac{n\pi}{2}\right)}{\cosh\left(\frac{n\pi c}{2b}\right)}$$

with $c = 1$ mm and $b = 50\,\mu$m. Since in our case the $z$ direction is much shorter than the others, this is the length that sets the boundary layer in the system. Therefore, the flow looks parabolic in the $z$-axis, but almost planar in the $y$-axis, with a very sharp drop to zero close to the boundaries.